\documentclass[journal]{IEEEtran}
\usepackage{graphicx}
\usepackage{amsmath}
\usepackage{tabularray}
\usepackage{color}
\usepackage{xcolor}
\usepackage{comment}
\newcommand{\ignore}[1]{}

\ifCLASSINFOpdf
\else
\fi
\begin{document}
%
\title{A 4$\times$32\,Gb/s 1.8\,pJ/bit Collaborative Baud-Rate CDR with Background Eye-Climbing Algorithm and Low-Power Global Clock Distribution}
%
%
%

\author{Jihee~Kim,~\IEEEmembership{Graduate~Student~Member,~IEEE,}
        Jia~Park,~\IEEEmembership{Graduate~Student~Member,~IEEE,}
        Jiwon~Shin,~\IEEEmembership{Member,~IEEE,}
        Hanseok~Kim,~\IEEEmembership{Graduate~Student~Member,~IEEE,}
        Kahyun~Kim,~\IEEEmembership{Graduate~Student~Member,~IEEE,}
        Haengbeom~Shin,~\IEEEmembership{Graduate~Student~Member,~IEEE,}
        Ha-Jung~Park,~\IEEEmembership{Graduate~Student~Member,~IEEE,}
        and~Woo-Seok~Choi,~\IEEEmembership{Member,~IEEE}
\thanks{This work was supported in part by Samsung Electronics Co., Ltd. (IO201221-08237-01) and by Institute of Information \& communications Technology Planning \& Evaluation (IITP) grant funded by the Korea government (MSIT) (No.2021-0-00871, Development of DRAM-Processing-In-Memory Chip for DNN Computing).}
\thanks{J. Kim, J. Park, J. Shin, H. Kim, K. Kim, H. Shin, H.-J. Park, and W.-S. Choi are with the Department of Electrical and Computer Engineering and the Inter University Semiconductor Research Center, Seoul National University, Seoul 08826, South Korea (email: jihee0548@snu.ac.kr, wooseokchoi@snu.ac.kr).}
\thanks{H. Kim is also with Samsung Electronics, Hwaseong 18448, South Korea.}}

%
%

\markboth{Journal of \LaTeX\ Class Files,~Vol.~14, No.~8, August~2015}%
{Shell \MakeLowercase{\textit{et al.}}: Bare Demo of IEEEtran.cls for IEEE Journals}
%



\maketitle

\begin{abstract}
This paper presents design techniques for an energy-efficient multi-lane receiver (RX) with baud-rate clock and data recovery (CDR), which is essential for high-throughput low-latency communication in high-performance computing systems. 
The proposed low-power global clock distribution not only significantly reduces power consumption across multi-lane RXs 
but is capable of compensating for the frequency offset without any phase interpolators.
To this end, a fractional divider controlled by CDR is placed close to the global phase locked loop.
Moreover, in order to address the sub-optimal lock point of conventional baud-rate phase detectors, the proposed CDR employs a background eye-climbing algorithm, which optimizes the sampling phase and maximizes the vertical eye margin (VEM). 
Fabricated in a 28\,nm CMOS process, the proposed 4$\times$32\,Gb/s RX shows a low integrated fractional spur of -40.4\,dBc at a 2500\,ppm frequency offset. 
Furthermore, it improves bit-error-rate performance by increasing the VEM by 17\,\%. 
The entire RX achieves the energy efficiency of 1.8\,pJ/bit with the aggregate data rate of 128\,Gb/s.
\end{abstract}


%
\IEEEpeerreviewmaketitle

\section{Introduction}
\begin{figure}
    \centering
    \includegraphics[width=1\linewidth]{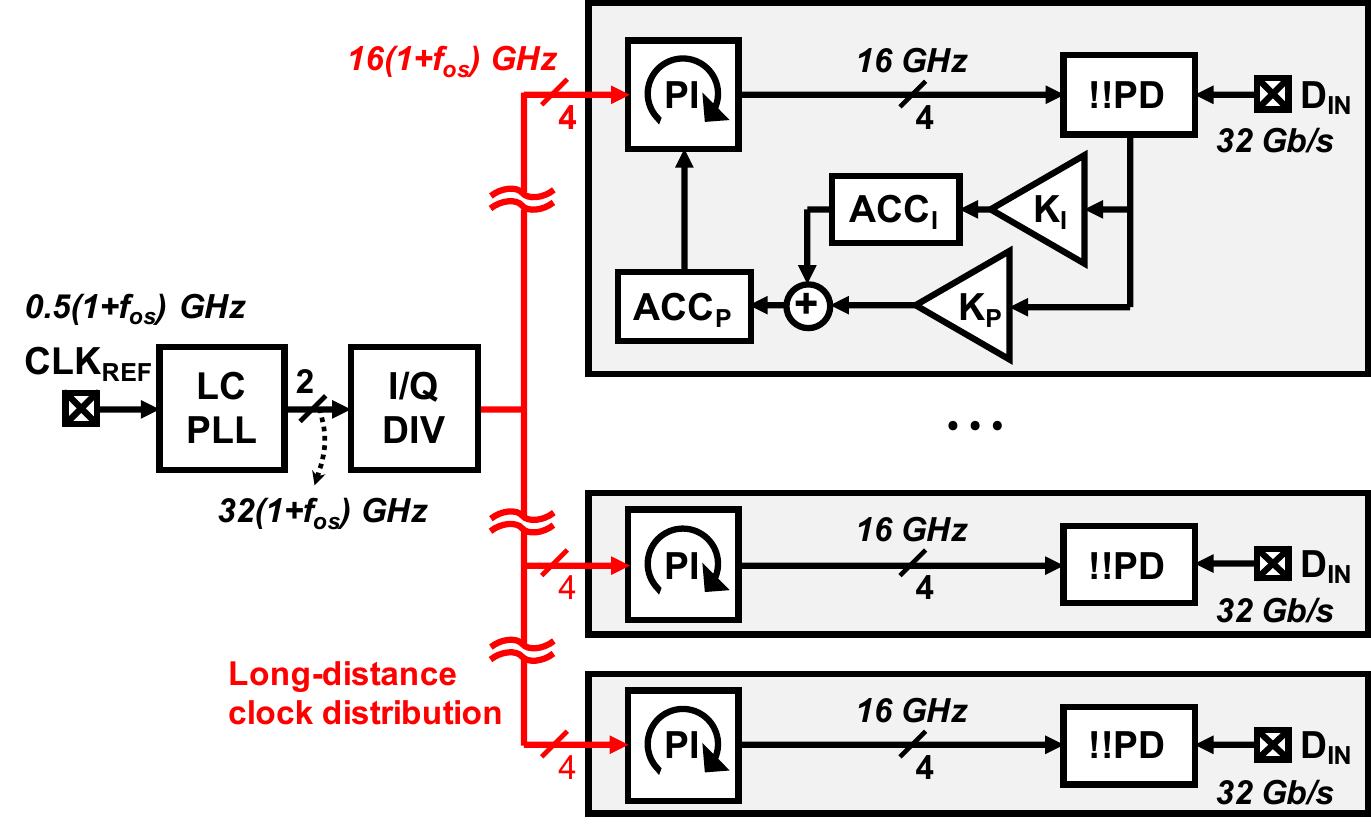}
    \caption{Clock distribution network of conventional PI-based multi-lane RX with a frequency error.}
    \label{fig: 1}
\end{figure}

%
%
%
%
\IEEEPARstart{M}{ulti-lane} high-speed wireline transceivers (TRXs) are essential building blocks in contemporary high-performance computing systems and data centers, enabling high-throughput low-latency communication. 
Such multi-lane TRXs have widely used phase interpolator (PI)-based clock and data recovery circuits (CDRs)~\cite{1}-\nocite{2,3,4,5,6}\cite{7}. 
Fig.~\ref{fig: 1} shows the conventional clock distribution network of PI-based CDRs in multi-lane 32\,Gb/s half-rate receivers (RXs). 
These CDRs distribute high-speed multi-phase clocks for the PI inputs from a global LC phase-locked loop (PLL)~\cite{1}-\nocite{2,3,4,5,6}\cite{7}. 
However, this structure has several drawbacks. 
First, it leads to excessive power consumption for clock distribution, which scales with clock frequency and the number of clock phases~\cite{low_power}. 
Second, it has limited frequency error tolerance as the PIs should track the frequency error between transmitted data and the RX-side reference clock~\cite{casper2009clocking}. 
Third, clock jitter performance is degraded due to the significant fractional spurs caused by the PI non-linearity and quadrature phase error generated during distribution.


\begin{figure}
    \centering
    \includegraphics[width=1\linewidth]{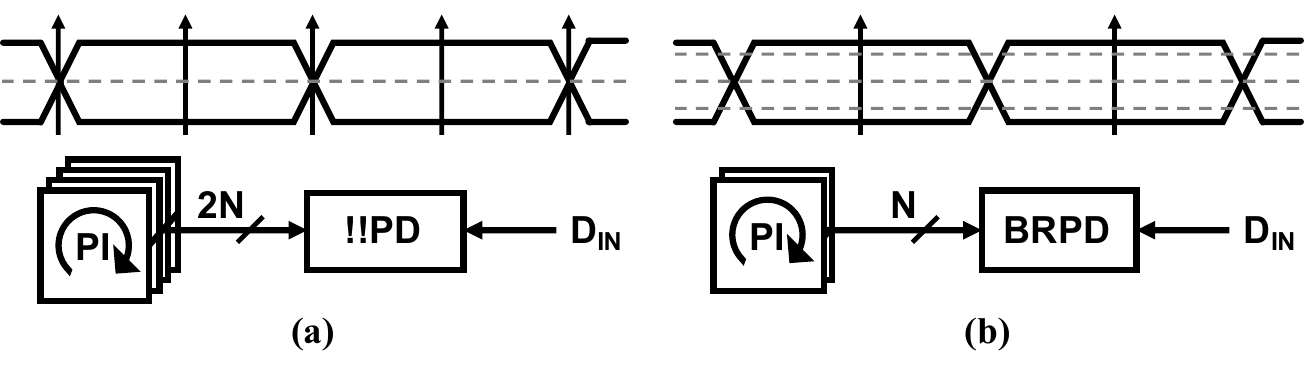}
    \caption{Comparison of sampling operations in (a) 2$\times$-oversampling CDR and (b) baud-rate CDR.}
    \label{fig: 2}
\end{figure}

\begin{figure}
    \centering
    \includegraphics[width=0.8\linewidth]{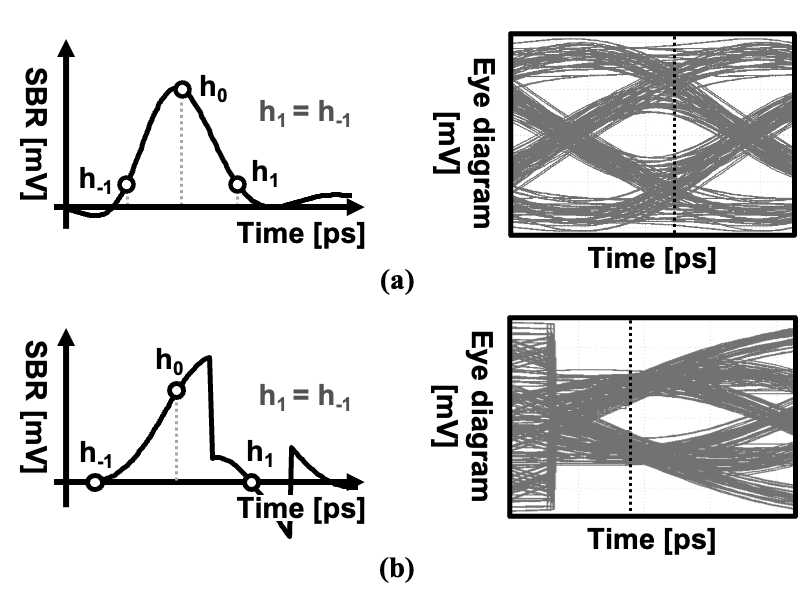}
    \caption{Lock point of conventional SS-MMPD (a) without DFE and (b) with DFE.}
    \label{fig: 3}
    \vspace{-1em}
\end{figure}

To reduce the clock distribution power, \cite{8} locates the PI near the LC PLL, reduces the clock frequency through a divider and distributes a low-speed single-phase clock to RX. 
The local RX ring-oscillator-based-PLL (RPLL), controlled by the CDR proportional path, then multiplies the frequency, generates multiple phases, and adjusts the sampling clock phases to recover data. 
While this approach reduces clock distribution power, it does not address the PI design issues since the PI should still track the frequency error. 
Furthermore, it leads to considerable power overhead to generate multi-phase clocks due to the use of bang-bang phase detectors (!!PDs) or 2$\times$-oversampling PDs. 
Fig.~\ref{fig: 2} compares the sampling operation of !!PD and baud-rate PD (BRPD). 
The necessity of 2$\times$ more samples per unit interval (UI) in !!CDRs doubles the number of PIs and sampling clock phases compared to BRCDRs~\cite{2,5}\nocite{6,7}-\cite{8}. 
Therefore, most clock recoveries in high-speed RXs rely on BRPDs for better energy efficiency~\cite{9}\nocite{10,11}-\cite{12}. 
However, the typical BRPD, sign-sign Mueller-Müller PD (SS-MMPD) faces challenges with its lock point, where $h_{1}$ and $h_{-1}$ coincide~\cite{18}-\nocite{19}\cite{20}. 
Fig.~\ref{fig: 3} shows the lock point of conventional SS-MMPD on single-bit-responses (SBRs) and eye diagrams, which results in reduced vertical eye margin (VEM) and increased bit-error-rate (BER). 
Moreover, with decision-feedback equalizers (DFEs), the lock point ($h_1=h_{-1}=0$) becomes susceptible to noise, consequently worsening the performance (see Fig.~\ref{fig: 3}(b)).

In view of these drawbacks, this paper introduces an energy-efficient multi-lane RX architecture employing a frequency-tracking, low-power global clock distribution network and a background eye-climbing algorithm (ECA). 
By distributing a low-frequency single-phase global clock and utilizing BRPDs, the proposed RX significantly reduces clocking power. 
The global clock frequency is adjusted by a fractional divider (FDIV), which is controlled by the CDR's integral path to track the frequency error, instead of PI. 
This approach enhances frequency error tolerance and minimizes fractional spurs. 
In addition, a background ECA effectively addresses the issues of the conventional MMPD and achieves the optimal lock point with the largest VEM. 

Rest of this paper is organized as follows. 
Clocking issues in conventional PI-based CDRs, followed by the proposed clock distribution network, is described in Section~\ref{section2}. 
Section~\ref{section3} explains the operation principle of the proposed ECA and baud-rate clock recovery method. 
Then, Section~\ref{section4} illustrates the overall architecture of the proposed multi-lane RX and its implementation details. 
The measurement results of the prototype chip are presented in Section~\ref{section5}. 
Finally, Section~\ref{section6} summarizes and concludes this paper.

\section{Clock Distribution for Multi-Lane Receiver} \label{section2}

\subsection{Limitations of Conventional PI-Based CDRs} \label{section_PI}
\begin{figure}
    \centering
    \includegraphics[width=1\linewidth]{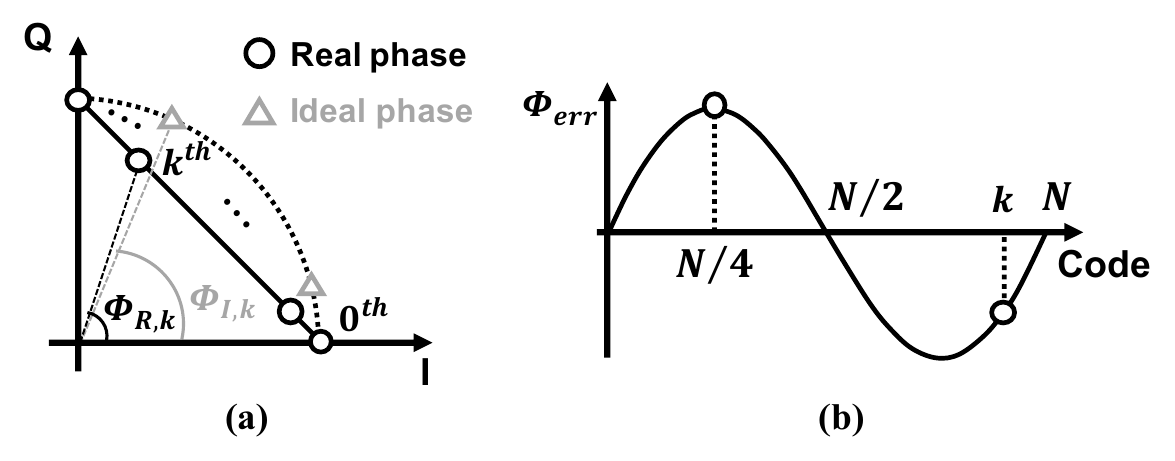}
    \caption{PI with uniform weights. (a) Phase constellation. (b) Phase error.}
    \label{fig: 4}
\end{figure}
\begin{figure}
    \centering
    \includegraphics[width=1\linewidth]{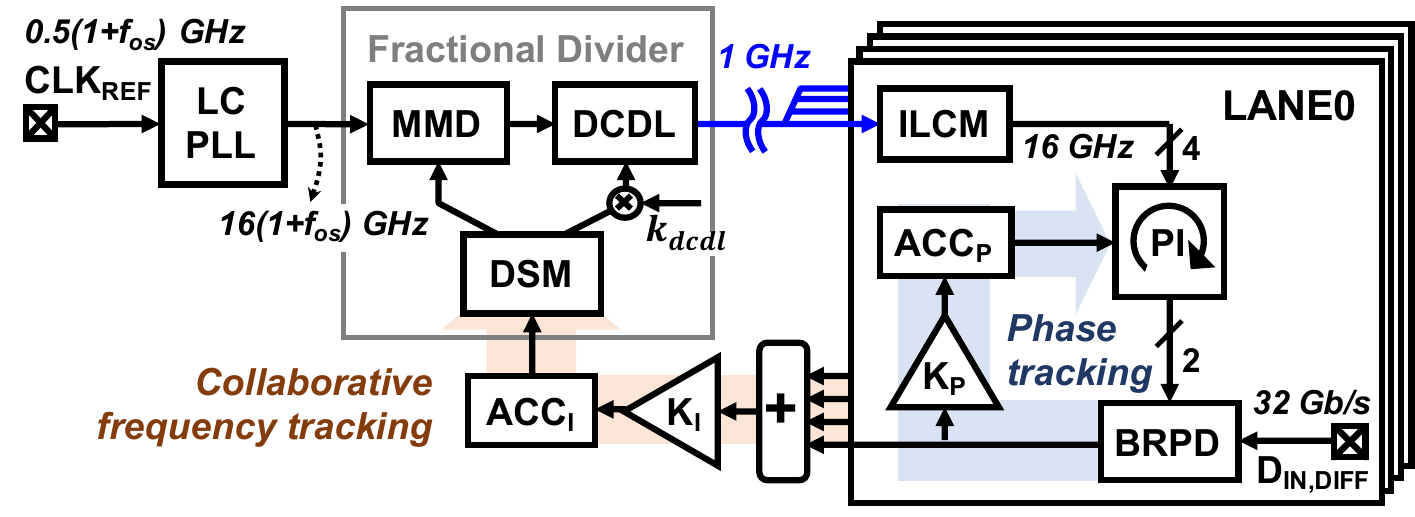}
    \caption{Overall architecture of the proposed clocking for multi-lane RX.}
    \label{fig: 5}
\end{figure}
In conventional multi-lane RXs with PI-based CDRs, high-frequency multi-phase clocks are typically distributed for PI inputs~\cite{casper2009clocking}. 
However, this method introduces design challenges associated with clock distribution power and PI non-linearity. 
First, distributing high-frequency multi-phase clocks incurs power consumption proportional to the number of phases and the clock frequency. 
Moreover, more clock buffers are required at a higher speed, causing additional power consumption. 
As the number of lanes grows, so does the distribution distance, leading to a further increase in power consumption. 
While high-frequency single-phase clock distribution can reduce some power demands, it requires additional power and hardware to locally generate multi-phase clocks, such as delay-locked loops (DLLs) or injection-locked oscillators (ILOs)~\cite{casper2009clocking}.

Second, the local PIs in conventional PI-based CDRs need to track both frequency and phase errors. 
In the presence of frequency error between TX and RX in separate clock domains, the PI's control code should rotate continuously. 
It reveals the PI's non-linearity as unwanted spurs, or deterministic jitter, even under ideal conditions with ideal sinusoidal inputs and uniform weighting. 
Fig.~\ref{fig: 4} shows the expected and actual output phases of PI that generates $N$ phases between quadrature clocks. 
The uniform interpolation of these sinusoidal waves forms a diamond-shaped constellation, not a uniform phase distribution, leading to sinusoidal phase errors, even in the absense of other non-linear factors such as input slew rate~\cite{13}. 
Furthermore, reducing this nonlinearity-induced jitter requires a finer resolution for the PI, leading to a trade-off between PI resolution and the ability to tolerate larger frequency error~\cite{15}.

Prior art such as \cite{rambus} tried to mitigate this problem by placing the frequency-tracking PI in the feedback path of the global PLL, where the phase error caused by the finite resolution and non-linearity of PI can be effectively reduced using the PLL's phase-domain low-pass characteristics. 
However, since this approach relies on PLL to suppress the PI-induced phase error, low PLL bandwidth is required, which conflicts with VCO phase noise suppression.


\subsection{Proposed Clock Distribution Network}

\begin{figure}
    \centering
    \includegraphics[width=1\linewidth]{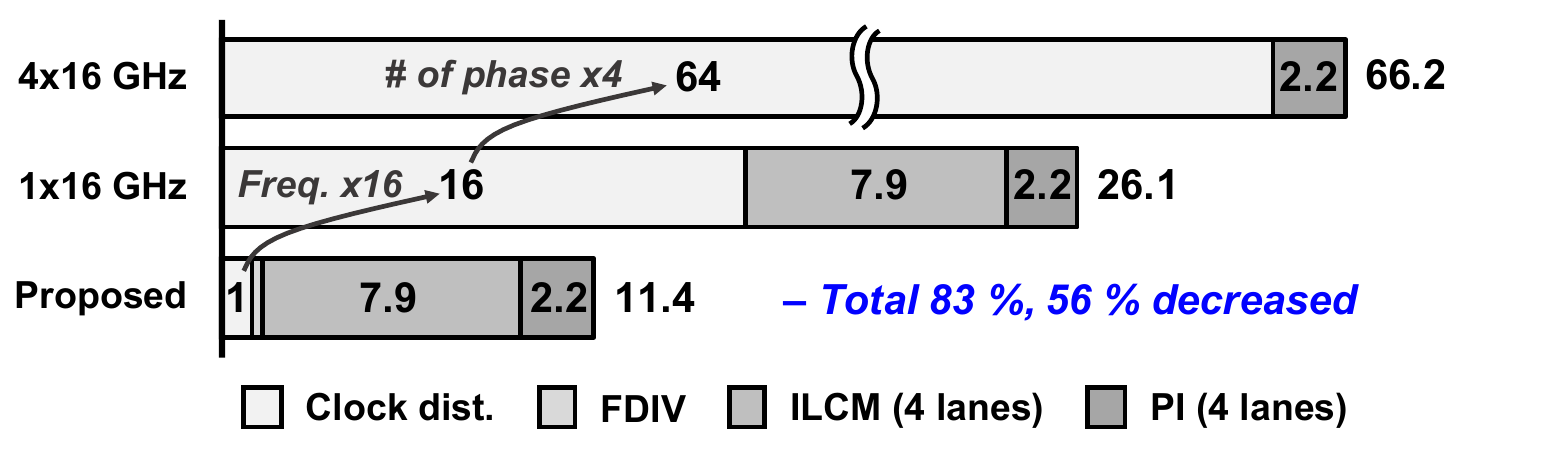}
    \caption{Comparison of clocking power estimated by post-layout simulation results.}
    \label{fig: 6}
\end{figure}
\begin{figure}
    \centering
    \includegraphics[width=1\linewidth]{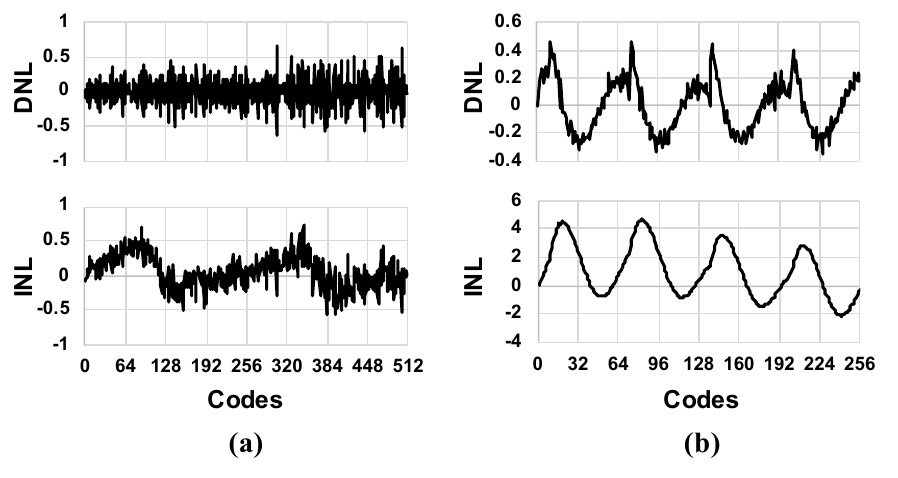}
    \caption{Simulated DNL and INL of (a) DCDL and (b) PI.}
    \label{fig: inl}
\end{figure}
Fig.~\ref{fig: 5} shows the overall structure of the proposed CDR with a focus on the clock distribution network. 
Instead of distributing multi-phase high-frequency (4$\times$16\,GHz) clocks or a single-phase 16\,GHz clock, the proposed CDR distributes a single-phase 1\,GHz clock to minimize the distribution power. 
Within each lane, an injection-locked clock multiplier (ILCM) locally generates 4-phase 16\,GHz clocks provided to local PIs for phase tracking. 
Power comparison, estimated based on post-layout simulation results, is presented in Fig.~\ref{fig: 6}. 
We assume that the identical PIs are used for all cases, and the same ILO is utilized for multi-phase generation in the CDR with single-phase clock distribution. 
In Fig.~\ref{fig: 6}, all the power values are normalized by the clock distribution power of the proposed CDR. 
Clocking power of the proposed RX is greatly reduced by 83\,\% and 56\,\%, respectively, 
which is attributed to the drastic reduction in global clock distribution power. 
We also note that, since single-phase high-frequency clock distribution requires additional hardware like ILO for multi-phase generation locally, distributing a low-frequency global clock and using local ILCMs that can multiply frequency as well as generate multi-phase clocks simultaneously with overhead similar to ILO, is more power-efficient.

In the proposed CDR, the FDIV, which consists of a multi-modulus divider (MMD), a gain-calibrated digitally-controlled delay line (DCDL), and a delta-sigma modulator (DSM)~\cite{16}, compensates for the frequency error, instead of PI.
The division ratio of the FDIV is controlled by accumulating the phase errors detected by the CDR logic, ensuring that local PIs get zero frequency offset clocks. 
This method provides two benefits: 1) the recovered clock shows lower deterministic jitter (i.e., fractional spur) due to the superior linearity of the DCDL in FDIV over PI, and 2) the trade-off between the PI resolution and its ability to track larger frequency error is alleviated, as the PI no longer tracks frequency error. 
In addition, the collaborative collection of error information across all lanes increases the transition density (phase error detection density) effectively, thereby allowing RX to track frequency error more accurately compared to conventional baud-rate CDRs~\cite{17}.
\begin{figure}
    \centering
    \includegraphics[width=0.8\linewidth]{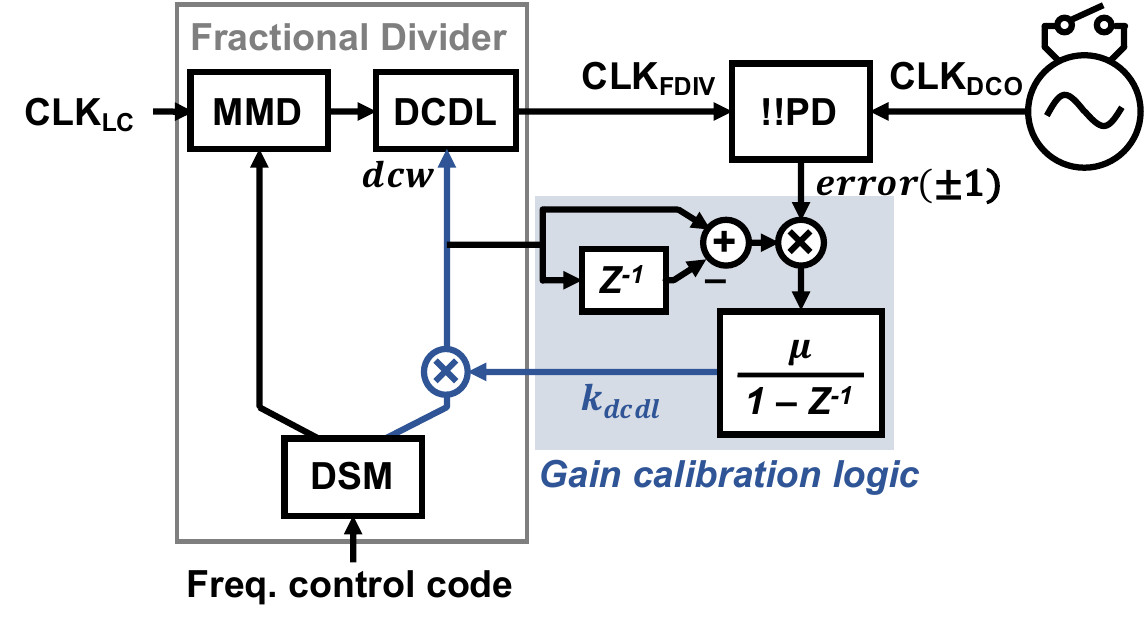}
    \caption{DCDL gain calibration with LMS.}
    \label{fig: 7}
\end{figure}
\begin{figure}
    \centering
    \includegraphics[width=0.6\linewidth]{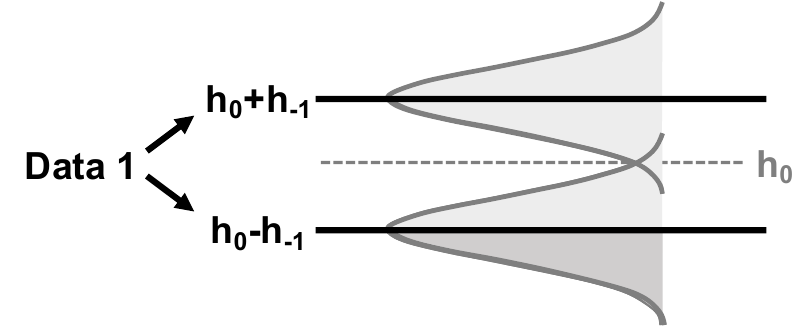}
    \caption{Data-level distribution with white Gaussian noise.}
    \label{fig: 8}
\end{figure}

Fig.~\ref{fig: inl} highlights the enhanced linearity of DCDL compared to PI by comparing their differential non-linearity (DNL) and integral non-linearity (INL). 
DCDL demonstrates maximum DNL and INL of 0.68\,LSB and 0.73\,LSB, respectively, while PI exhibits 0.46\,LSB and 4.67\,LSB, respectively. 
Notably, DCDL achieves better linearity despite its finer phase step of 183\,fs, compared to the PI's of 244\,fs.
Designing linear phase rotators has been a difficult task as discussed in Section~\ref{section_PI}. 
Although prior art~\cite{PR1}-\nocite{PR2, 8b}\cite{7b} has introduced effective phase rotators, they exhibit average DNL and INL of 0.68\,LSB and 1.08\,LSB, respectively, similar to or worse than those of DCDL, even with an average phase step of 719\,fs, 
which emphasizes superior linearity of DCDL.

To minimize the phase error caused by FDIV, the range of DCDL must be equivalent to the period of the FDIV's input clock ($\textrm{CLK}_\textrm{LC}$) under process, voltage, and temperature (PVT) variations~\cite{7027236}.
Consequently, background calibration of the DCDL gain ($k_{dcdl}$) is necessary. 
The calibration process employs the sign-sign least-mean-square (SS-LMS) algorithm, correlating the bang-bang PD (!!PD) output from ILCM with changes in the DCDL control code ($dcw$), as illustrated in Fig.~\ref{fig: 7}. 
Assuming the constant period of $\textrm{CLK}_\textrm{FDIV}$ (ILCM input), the error from !!PD can be expressed as:
\begin{equation}
\begin{aligned}
    error &= (T_{mmd}[n]+k_{dcdl}\cdot{}dcw[n]) \\&-(T_{mmd}[n+1]+k_{dcdl}\cdot{}dcw[n-1]), 
\end{aligned}
\end{equation}
where $T_{mmd}$ represents the period of the MMD output clock. Applying this error to the LMS algorithm leads to:
\begin{equation}
    k_{dcdl}[n+1]=k_{dcdl}[n]-\mu\nabla_{k}error^{2}
\end{equation}
\begin{equation}
    k_{dcdl}[n+1]=k_{dcdl}[n]-\mu\cdot2error\cdot(dcw[n]-dcw[n-1]).
\end{equation}
By correlating the difference between the current $dcw$ and the previous $dcw$ with the error, it enables determination of an appropriate $k_{dcdl}$ and ensures precise FDIV operation.
\begin{figure}
    \centering
    \includegraphics[width=1\linewidth]{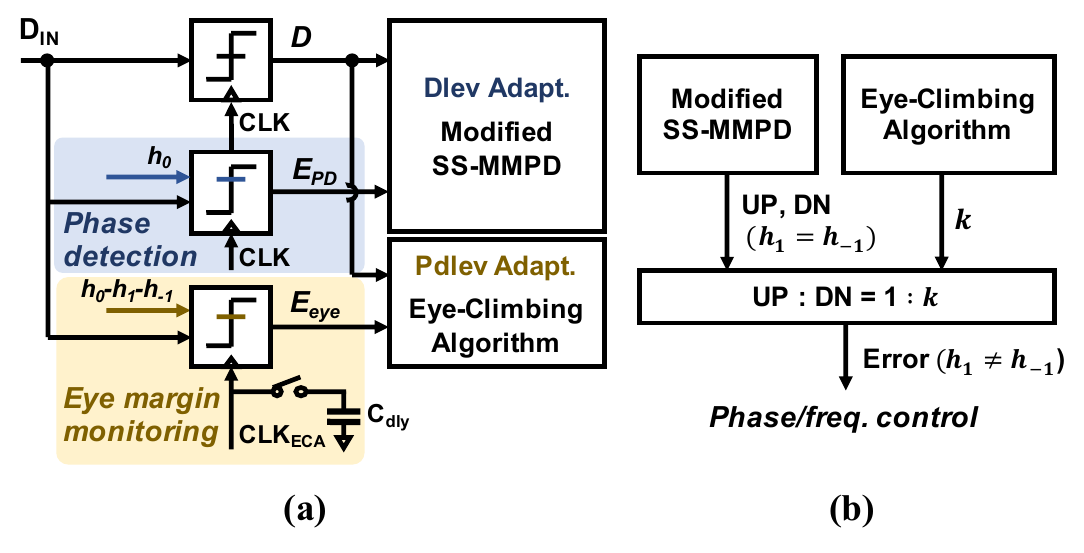}
    \caption{(a) Three samplers for proposed CDR. (b) Phase update process.}
    \label{fig: 9}
\end{figure}
\begin{figure}
    \centering
    \includegraphics[width=0.95\linewidth]{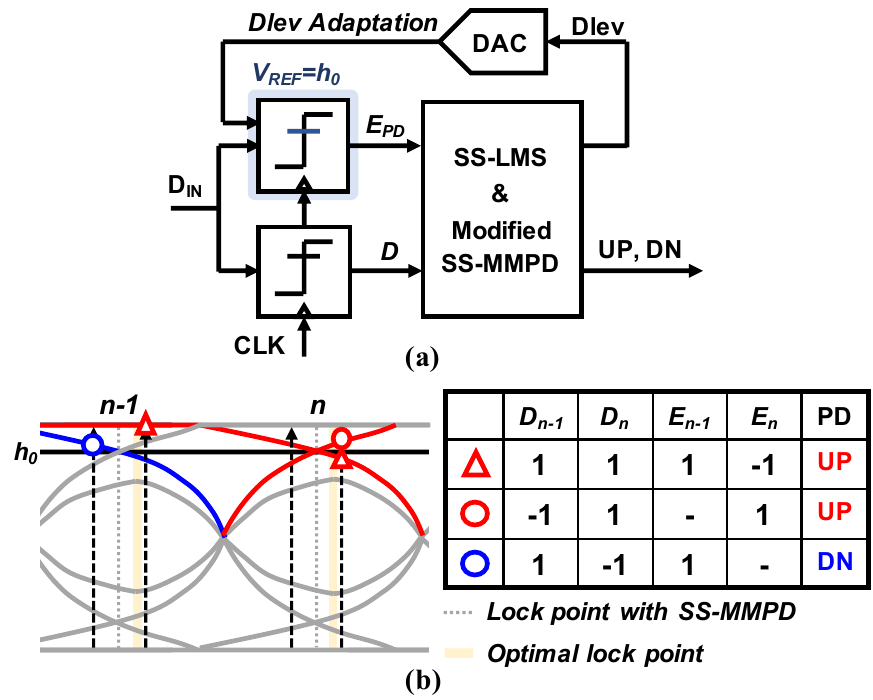}
    \caption{(a) Phase detection in proposed CDR. (b) Phase detection logic.}
    \label{fig: 11}
\end{figure}


\section{Clock and Data Recovery} \label{section3}
\subsection{Issues in Prior Baud-Rate CDRs}
Prior art has tried to address the issue of the sub-optimal lock point in conventional SS-MMPD-based CDRs~\cite{18}. 
For instance, \cite{19} proposed an unequalized MMCDR that intentionally adds offset to enable phase lock at a point where $h_1$ and $h_{-1}$ do not coincide. 
Choosing a proper offset results in enhanced voltage margin, thereby improving BER. 
However, manually finding the optimal offset value, which is sensitive to channel characteristics, makes it difficult to use \cite{19} in practice.
\begin{figure}
    \centering
    \includegraphics[width=0.7\linewidth]{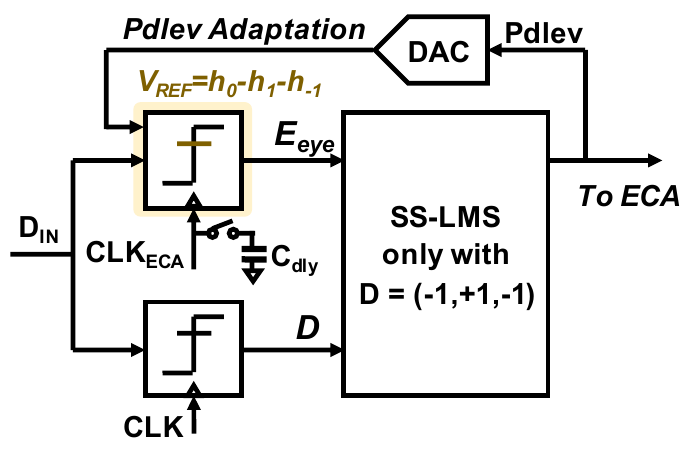}
    \caption{Eye margin monitoring through Pdlev adaptation.}
    \label{fig: 12}
\end{figure}
\begin{figure}
    \centering
    \includegraphics[width=0.9\linewidth]{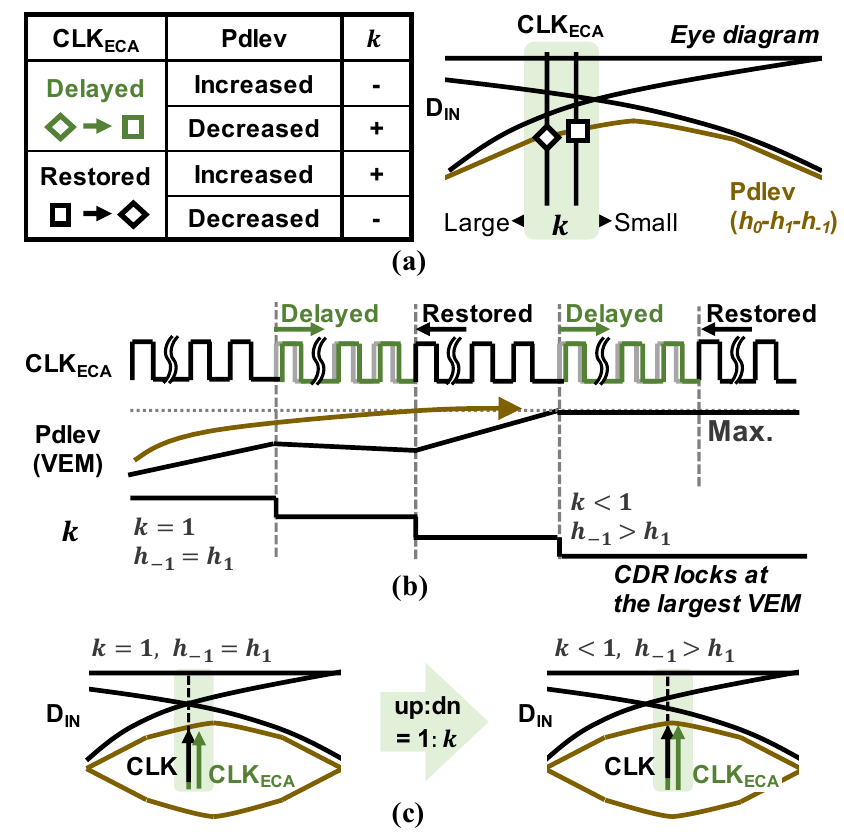}
    \caption{(a) Operation principle and (b) transient example of proposed ECA. (c) Change of sampling clock phase.}
    \label{fig: 13}
\end{figure}

\begin{figure*}
    \centering
    \includegraphics[width=0.8\linewidth]{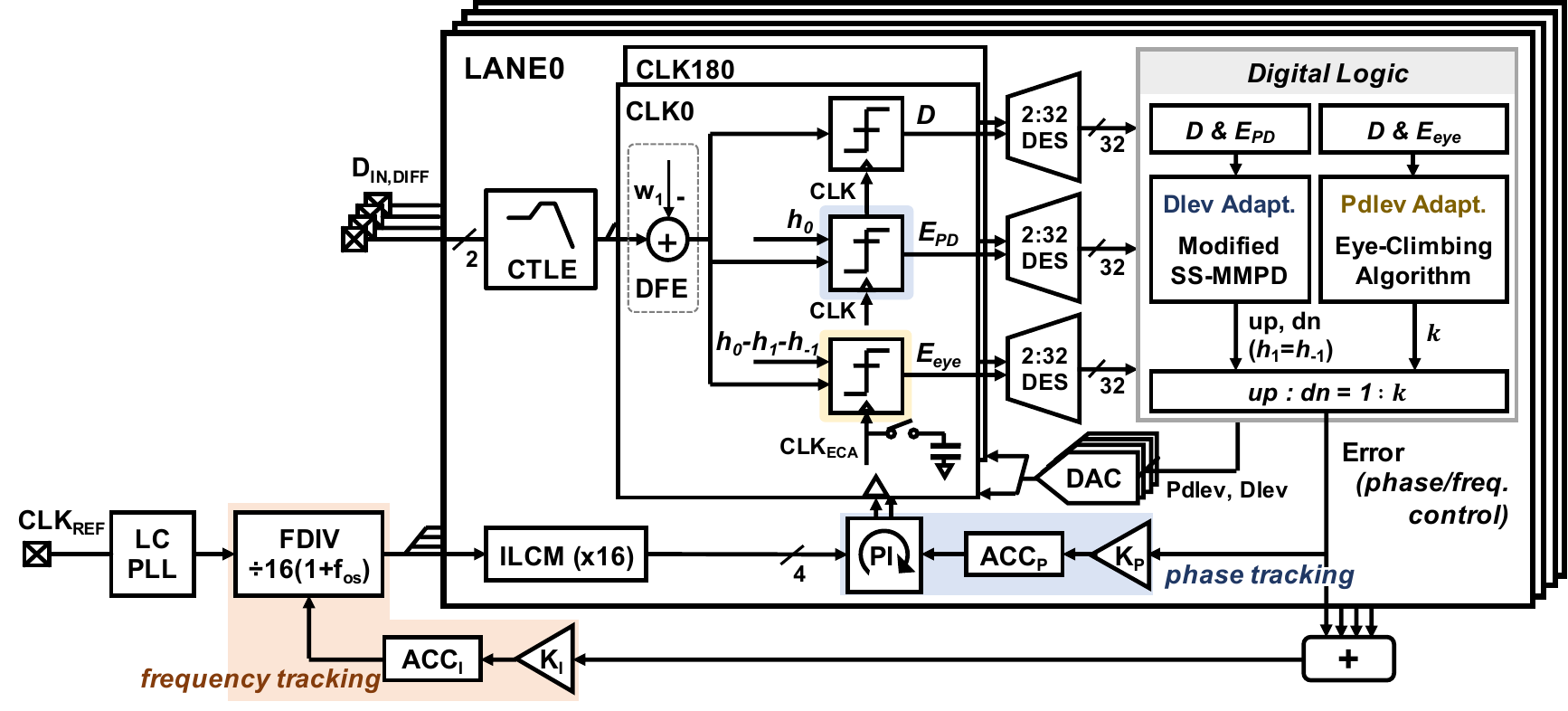}
    \caption{Overall architecture of the proposed 4-lane CDR.}
    \label{fig: archi}
\end{figure*}

\begin{figure*}
    \centering
    \includegraphics[width=0.9\linewidth]{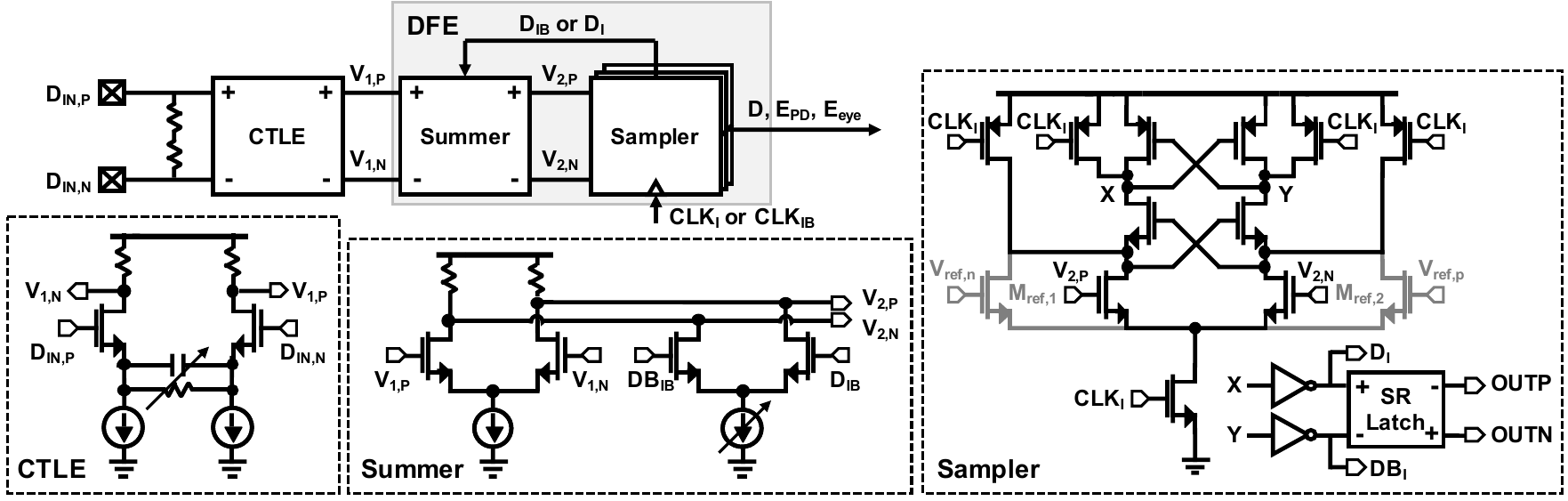}
    \caption{Analog front-end implementation.}
    \label{fig: 17}
\end{figure*}
To mitigate this issue, \cite{20} automatically search for the optimal offset by monitoring the eye height. 
When DFE removes all post-cursors and only the first precursor remains, data 1 can have two voltage levels, $h_0+h_{-1}$ and $h_0-h_{-1}$, and the eye height can be represented as $h_0-h_{-1}$ (see Fig.~\ref{fig: 8}). 
Then, by updating the data-level with LMS in 1:3 ratio, the adapted data-level converges to $h_0-h_{-1}$, effectively indicating the eye height~\cite{MET}. 
While sweeping the offset, this biased data-level (Bdlev) adaptation can be utilized to identify the eye height corresponding to each offset and choose the one with the maximum eye height. 
However, this CDR also has several drawbacks. 
First, operating with only two samplers reduces the transition density to half that of the conventional SS-MMPD. 
Besides, using a single error sampler for both phase detection and Bdlev adaptation for eye height monitoring can cause interaction between two loops, resulting in undesirable dead zones. 
Second, when noise power is comparable to $h_{-1}$, the overlap region between the two levels increases as depicted in Fig.~\ref{fig: 8}, which makes the Bdlev adaptation inadequate for accurately representing the effective eye height. 
Finally, while this method allows for automatic offset determination, it cannot operate robustly against PVT variations, necessitating a background calibration technique.


\subsection{Proposed Baud-Rate CDR with ECA}
The proposed baud-rate CDR operates with three samples per UI as shown in Fig.~\ref{fig: 9}(a): One for data recovery, a second for SS-MM phase detection, and a third for eye margin monitoring. 
The phase detecting error sampler’s outputs are used for data-level (Dlev) adaptation and modified SS-MMPD, 
and the outputs of the eye monitoring sampler are for pattern-based data-level (Pdlev) adaptation indicating the eye margin. 
Fig.~\ref{fig: 9}(b) shows the phase update process of the proposed CDR. 
Initially, CDR locks at the phase where $h_1=h_{-1}$ with modified SS-MMPD, employing only one error sampler. 
Subsequently, the clock phase update weight is changed to $up:dn=1:k$ to adjust the recovered clock phase. 
The ratio or offset $k$ is calibrated by the proposed ECA to ensure CDR locks at the largest VEM, maximizing Pdlev. 
Note that the clock phase ($\textrm{CLK}_{\textrm{ECA}}$) for the eye margin monitoring sampler differs from CLK used in the other two samplers. 
Depending on whether the switch is on or not, $\textrm{CLK}_{\textrm{ECA}}$ is either delayed or identical to CLK.

The proposed BRPD operates with just one error sampler and a data sampler but achieves higher transition density compared to other BRPDs with two samplers~\cite{20,56g,28}.
Fig.~\ref{fig: 11} demonstrates the operation of the modified SS-MMPD. 
The outputs from the error sampler ($E_{PD}$) and the data sampler ($D$) are utilized for phase error detection and Dlev adaptation. 
Dlev for the phase detection error sampler is adapted to $h_0$ with the SS-LMS algorithm. 
Fig.~\ref{fig: 11}(b) details the modification in the PD logic. 
By combining the error detection pattern from \cite{18} and \cite{20}, it allows for phase error detection in three cases with just two samplers, enhancing the transition density compared to conventional BRPDs with two samplers\footnote{Since the DFE removes post-cursors, the pattern ($D_{n-1},D_n,E_{n-1},E_n)$ = $(1,1,-1,1)$ does not occur. Thus, there are more patterns for detecting $up$ than for detecting $dn$.}.


For eye height estimation, the eye monitoring sampler's reference voltage (Pdlev) is adapted through SS-LMS, updated only with the data pattern $(-1,+1,-1)$, as shown in Fig.~\ref{fig: 12}. 
This makes Pdlev converge to $h_0-h_1-h_{-1}$, which represents the effective VEM. 
The input of the samplers, $\textrm{D}_{\textrm{IN}}$, can be expressed as convolution between data $x_n$ and channel's impulse response $h_n$:
\begin{equation}
\begin{aligned}
    \textrm{D}_{\textrm{IN},n}&=\sum(x_{n-k}h_k) \\
    &=\cdots+h_{1}x_{n-1}+h_0x_n+h_{-1}x_{n+1}+\cdots.   
\end{aligned}
\end{equation}
The SS-LMS algorithm applied with the pattern of $(-1,1,-1)$ finds Pdlev ($w_0$), which minimizes the expression: 
\begin{equation}
\begin{aligned}
    &\sum_n(\textrm{D}_{\textrm{IN},n}-w_0x_n)^2\\ 
    &=(\cdots+h_1x_{n-1}+h_0x_n+h_{-1}x_{n+1}+\cdots-w_0x_n)^2\\ 
    &=(\cdots-h_1+h_0-h_{-1}+\cdots-w_0)^2.
\end{aligned}
\end{equation}
Assuming that the input data is random, $w_0$, or Pdlev, converges to $h_0-h_1-h_{-1}$. 
As a result, Pdlev reliably represents VEM, unaffected by noise levels.
To accurately account for the residual $h_1$ due to the quantized DFE tap implementation, Pdlev is adapted to $h_0-h_1-h_{-1}$, not $h_0-h_{-1}$.

Fig.~\ref{fig: 13} details the operation principle and a transient example of the proposed ECA. 
Using the Pdlev, the proposed CDR “climbs” toward the top of the eye. Starting from an initial lock point $h_1=h_{-1}$ (i.e., $k=1$), the CDR then periodically dithers $\textrm{CLK}_{\textrm{ECA}}$ by turning on and off the delay cap, $\textrm{C}_{\textrm{dly}}$. 
Observing whether Pdlev is increased or not, it adjusts k accordingly. 
In the case of Fig.~\ref{fig: 13}, with an initially positive eye slope, Pdlev increases when $\textrm{CLK}_{\textrm{ECA}}$ is delayed ($\textrm{C}_{\textrm{dly}}$ on) and decreases when the $\textrm{CLK}_{\textrm{ECA}}$ phase is restored back ($\textrm{C}_{\textrm{dly}}$ off). 
Hence, in this case, decreasing $k$ shifts the clock phase rightward, achieving a larger VEM than with $k=1$. This process continues until CLK and $\textrm{CLK}_{\textrm{ECA}}$ settle to the point where Pdlev (or VEM) becomes maximum, as in Fig.~\ref{fig: 13}(c) right. 
Compared to the conventional SS-MMPD, this method enables RX to achieve a lower BER by securing the largest VEM without additional hardware. 
Moreover, even when the sampling clock reaches the largest VEM, the ECA continues dithering $\textrm{CLK}_{\textrm{ECA}}$ and monitoring eye margin. 
This background operation ensures the optimal lock point is maintained even with PVT variations.
\begin{figure}
    \centering
    \includegraphics[width=0.9\linewidth]{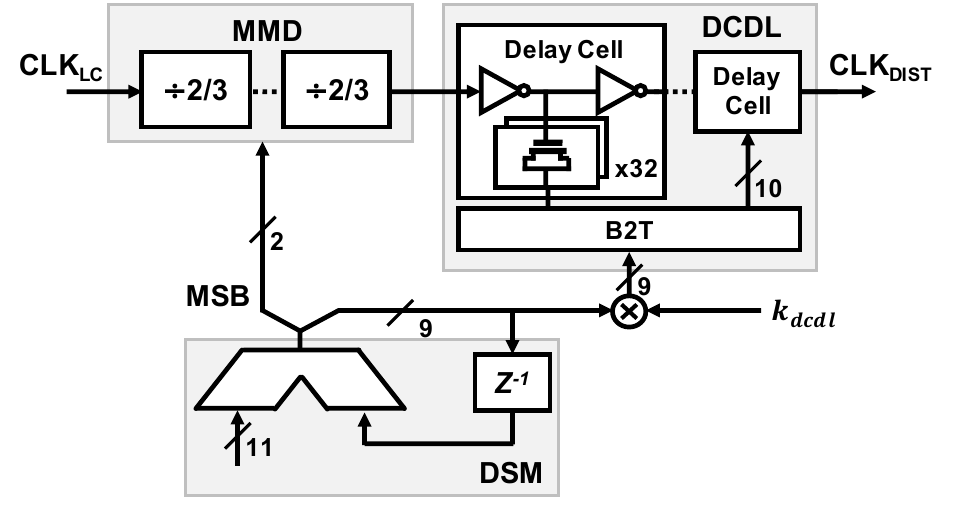}
    \caption{Fractional divider implementation.}
    \label{fig: 14}
\end{figure}
\begin{figure}
    \centering
    \includegraphics[width=0.7\linewidth]{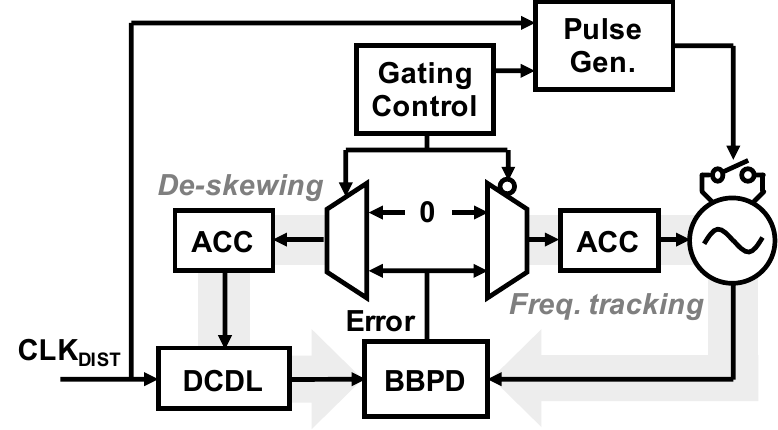}
    \caption{ILCM architecture.}
    \label{fig: 15}
\end{figure}

\section{Implementation Details} \label{section4}

Fig.~\ref{fig: archi} shows the overall architecture of the proposed 4-lane RX. 
FDIV divides the LC PLL clock into a 1\,GHz clock and distributes it to each lane. 
The separated integral and proportional paths of CDR lead to improved frequency error tolerance and jitter performance. 
Utilizing the low-frequency single-phase distributed clock, each lane generates high-frequency multi-phase clocks with an ILCM, 
and PIs, taking the ILCM outputs as input, are controlled by the CDR proportional path and adjust the recovered clock phase. 
In the RX analog front-end (AFE), a continuous-time linear equalizer (CTLE) and a 1-tap DFE are employed to compensate for channel loss. 
Reference voltages for phase detection and eye margin monitoring samplers are generated using 6-bit resistor DACs according to the Dlev and Pdlev codes. 
The data and error samples are deserialized and processed by the digital block, operating at a 32$\times$ lower frequency.
\begin{figure}
    \centering
    \includegraphics[width=1\linewidth]{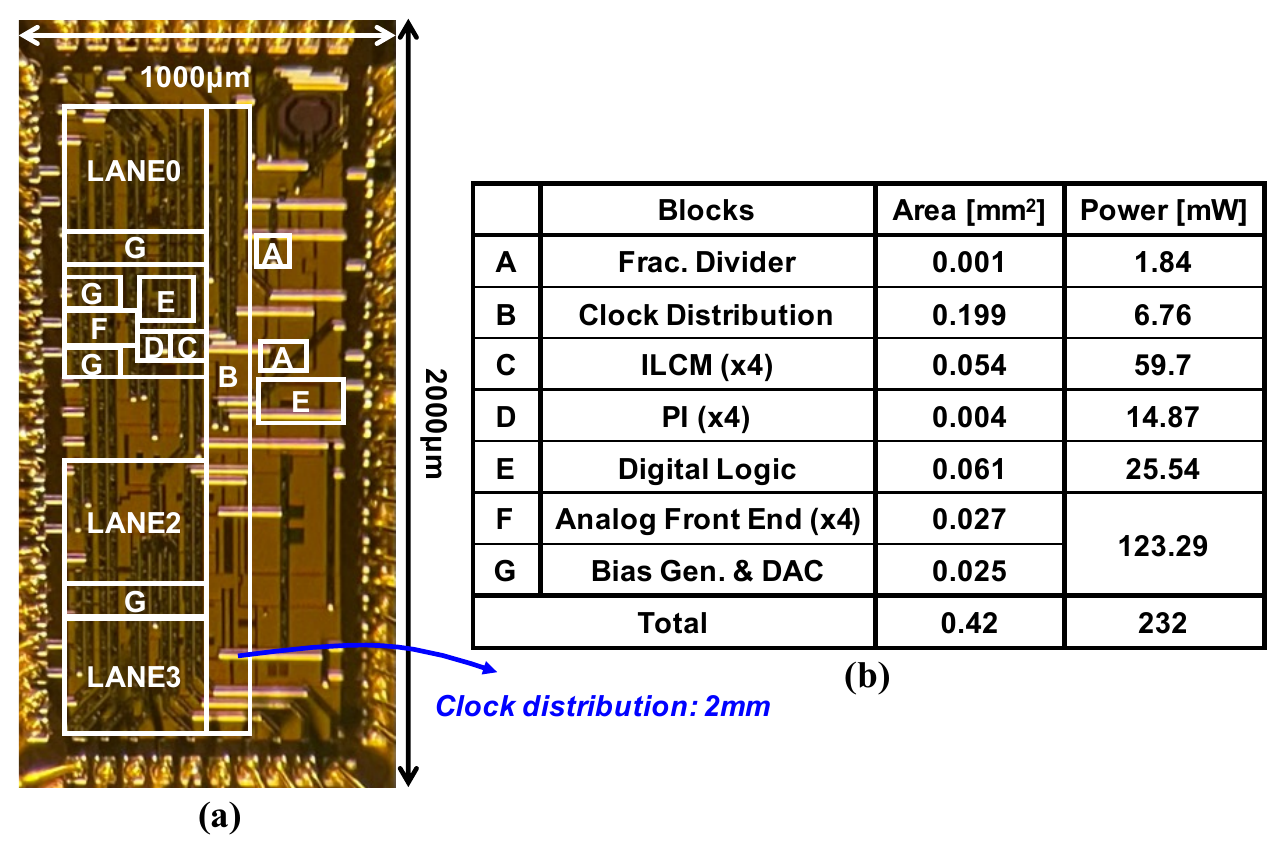}
    \caption{(a) Chip photomicrograph. (b) Active area and power breakdown.}
    \label{fig: 18}
\end{figure}
\subsection{Analog Front-End}
To compensate for a 15\,dB channel loss, RX incorporates a CTLE and a 1-tap DFE. 
The detailed schematic of the implemented AFE is depicted in Fig.~\ref{fig: 17}. 
The CTLE is designed to provide up to 8.9\,dB peaking with its degeneration capacitance and resistance being manually controllable~\cite{23}. 
The 1-tap DFE is implemented using a current-based summer to eliminate the first post-cursor~\cite{24}. 
The DFE summer output is connected to one data sampler and two error samplers, comprising strong-arm latches and SR latches~\cite{25}. 
Note that the error samplers incorporate additional transistors, $\textrm{M}_{\textrm{ref,1}}$ and $\textrm{M}_{\textrm{ref,2}}$, to provide offset for reference voltage comparison.

\subsection{Fractional Divider}
The implemented fractional divider, as illustrated in Fig.~\ref{fig: 14}, consists of an MMD, a DCDL, and a DSM. 
The DSM takes a frequency control code from the CDR integral path and generates the control codes for the MMD and DCDL.
The MMD, capable of seamless switching, has four divide-by-2/3 cells to enable a division range from 8 to 32~\cite{21}. 
The DCDL delay is controlled by switching MOS capacitors on/off. 
It comprises 16 delay cell stages in total, where each delay cell is controlled by 32 MOS capacitors. 
The 9-bit binary delay control code is converted into a mix of thermometer and binary code for layout simplicity~\cite{8357042}. 
To provide precise delay robustly against PVT variations, the DCDL control code is generated after multiplying a LMS-calibrated gain, $k_{dcdl}$.
The FDIV output CLK$_\textrm{DIST}$ is distributed to four local RX CDRs.
\begin{figure}
    \centering
    \includegraphics[width=0.9\linewidth]{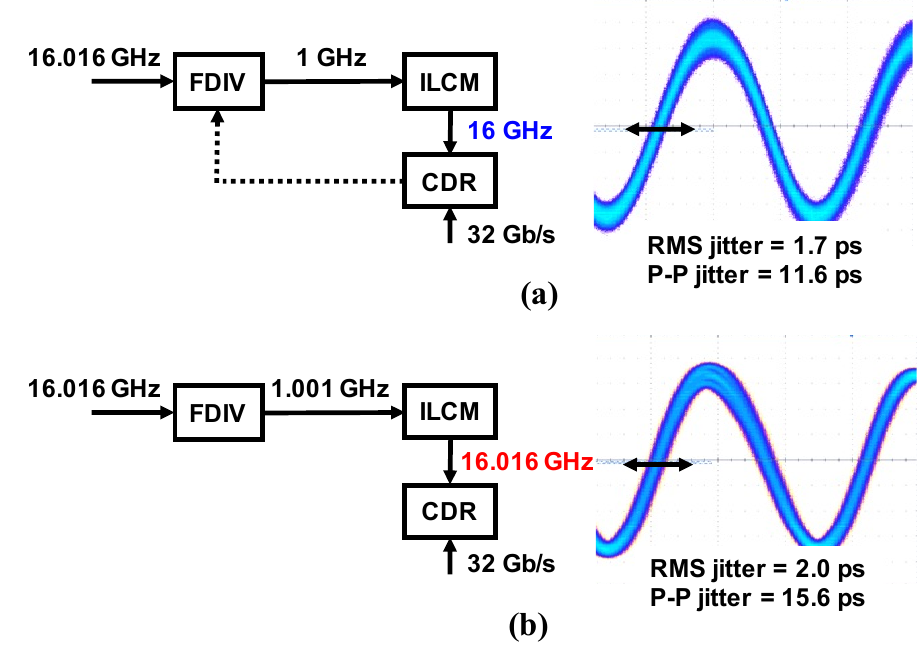}
    \caption{Measured recovered clock jitter with 1000\,ppm frequency offset: (a) Frequency tracking with FDIV. (b) Frequency tracking only with PI.}
    \label{fig: 20}
\end{figure}
\begin{figure}
    \centering
    \includegraphics[width=1\linewidth]{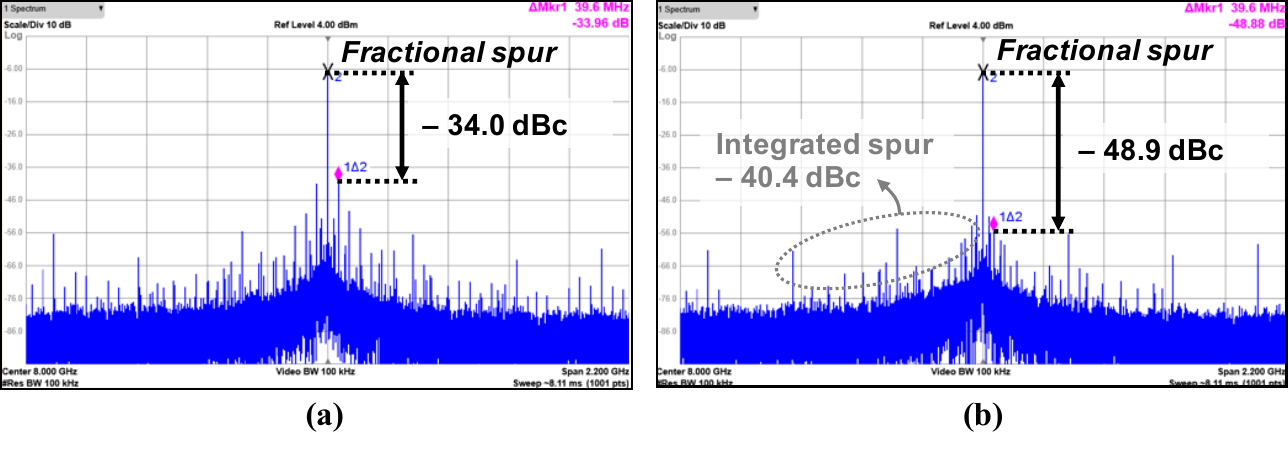}
    \caption{Measured fractional spur of ILCM output with 2500\,ppm frequency offset (a) before DCDL gain calibration and (b) after DCDL gain calibration.}
    \label{fig: 21}
\end{figure}
\subsection{Local Clock Path}
In each lane, the ILCM converts the distributed low-frequency clock into high-frequency multi-phase clocks, and the local PIs adjust the phase to generate the recovered clock. 
Fig.~\ref{fig: 15} details the structure of ILCM. 
As in \cite{22}, ILCM employs the gating approach to track the frequency of the reference clock and address delay mismatch through a de-skewing loop, alleviating reference spur. 
However, the de-skewing loop in the ILCM operates when the injection pulse is gated, differing from \cite{22} where the frequency tracking loop operates during pulse injection. 
For phase tracking, 8-bit current-mode PIs are implemented with seamless switching~\cite{7999180} and controlled by the CDR proportional path.


\section{Measurement Results} \label{section5}
The prototype 4-lane RX is fabricated in a 28\,nm CMOS and occupies an active area of 0.42\,mm$^2$ as shown in Fig.~\ref{fig: 18}. 
The entire RX consumes 232\,mW (58\,mW per lane) with the aggregate data rate of 128\,Gb/s. 
Fig.~\ref{fig: 18}(b) shows the area and power breakdown of the entire RX.


\subsection{Clocking Performance}
The proposed multi-lane RX efficiently distributes a low-frequency single-phase clock over a 2\,mm distance, consuming only 6.76\,mW. 
Moreover, thanks to the good DCDL linearity in FDIV, the recovered clock shows better jitter performance than PI-based CDR in the presence of frequency offset, 
which is demonstrated in Fig.~\ref{fig: 20} (measured with a 1000\,ppm frequency offset). 
As shown in Fig.~\ref{fig: 20}(a), the proposed CDR, where the FDIV tracks frequency offset and the PIs only cover phase error, shows the recovered clock jitter of 1.7\,ps$_\textrm{rms}$ and 11.6\,ps$_\textrm{p2p}$. 
On the other hand, if the frequency offset is tracked only by PIs as in conventional CDRs\footnote{To mimic the PI behavior in conventional CDRs, the control code for FDIV is fixed as shown in Fig~\ref{fig: 20}(b).}, jitter increase to 2.0\,ps$_\textrm{rms}$ and 15.6\,ps$_\textrm{p2p}$ (see Fig.~\ref{fig: 20}(b)).
Fig.~\ref{fig: 21} shows the measured spectra of the ILCM output with a 2500\,ppm frequency offset (FDIV division ratio fixed at 16.04), where fractional spurs due to DCDL non-linearity can be observed. 
Without DCDL gain calibration, fractional spur of -34.0\,dBc appears at 40\,MHz due to large phase-domain quantization error. 
However, with the LMS-based gain calibration, this spur is reduced to -48.9\,dBc. 
Table~\ref{tab: 1} compares the performance of the proposed frequency tracking scheme with other state-of-the-art multi-phase generating phase rotators~\cite{8b,7b}. 
Thanks to the superior DCDL linearity in FDIV, the lowest integrated fractional spur is achieved with excellent clocking power efficiency of 1\,mW/GHz.
\begin{table}
\centering
\caption{Performance comparison with other multi-phase generating phase rotators. }
\begin{tabular}{|c|c|c|c|}
\hline
  & ISSCC'19~\cite{8b} & ISSCC'21~\cite{7b} & \textbf{This Work} \\
\hline
Technology & 16\,nm & 65\,nm & 28\,nm \\ 
\hline
Architecture & ILPR& MPILOSC& FDIV+ ILCM \\ 
\hline
Resolution [bits]  & 8   & 7 & \textbf{9}    \\ 
\hline
Power [mW] & \begin{tabular}[c]{@{}c@{}}11.4\\@ 7\,GHz\end{tabular} & \begin{tabular}[c]{@{}c@{}}15.6\\@ 7\,GHz\end{tabular} & \begin{tabular}[c]{@{}c@{}}16.0\\@ 16\,GHz\end{tabular} \\ 
\hline
\begin{tabular}[c]{@{}c@{}}Power/GHz\\{[}mW/GHz]\end{tabular}  & 1.63  & 2.23  & \textbf{1} \\ 
\hline
Integrated & -39.4\,dBc & -33.9\,dBc & \textbf{-40.4\,dBc} \\
Fractional Spurs & @ -1300\,ppm & @ 1000\,ppm & @ 2500\,ppm \\
\hline

\end{tabular}
\label{tab: 1}
\end{table}

\begin{figure}
    \centering
    \includegraphics[width=0.7\linewidth]{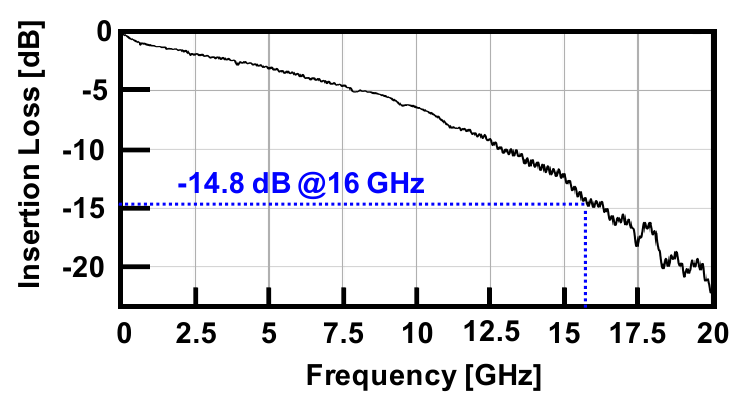}
    \caption{Measured insertion loss of the channel.}
    \label{fig: 22}
\end{figure}

\begin{table*}
\centering
\caption{Performance summary and comparison with high-speed NRZ RXs.}
\begin{tblr}{
 cells = {c},
  row{15} = {r},
  cell{3}{1} = {r=2}{},
  cell{10}{1} = {r=2}{},
  cell{15}{1} = {c=7}{},
  vline{-} = {1-14}{},
  vline{1,8} = {3,10}{},
  vline{2-7} = {3,10}{},
  vline{3-7} = {4,11}{},
  vline{8} = {4,11}{},
  hline{1-3,5-8,10,12-15} = {-}{},
  hline{9} = {-}{},
}
                                      & ISSCC’15~\cite{19} & VLSI’20~\cite{20} & ISSCC’17~\cite{16} & ISSCC’16~\cite{28} & JSSC’23~\cite{29} & \textbf{This work} \\
Technology                            & 14\,nm                                               & 28\,nm                                              & 65\,nm                                               & 28\,nm                                               & 22\,nm                                              & 28\,nm     \\
Phase Detection                       & Unequalized                                         & SS-MMPD                                            & 2x-                                                 & Pattern Based                                       & 2x-                                                & \textbf{SS-MMPD}   \\
                                      & SS-MMPD                                             & with MET                                           & Oversampling                                        & BRPD                                                & Oversampling                                       & \textbf{with ECA}  \\
Clock Recovery                        & PI*                                                  & PI*                                                 & FDIV + MDLL                                         & VCO                                                 & PLL + PI*                                           & FDIV +~PI* \\
\# of Lanes                           & 2                                                   & 1                                                  & 1                                                   & 2                                                   & 1                                                  & \textbf{4}         \\
Data Rate [Gb/s]                      & 10                                                  & 28                                                 & 10                                                  & 56.2                                                & 26                                                 & 32        \\
Aggregate Data~Rate [Gb/s]            & 20                                                  & 28                                                 & 10                                                  & 112.4                                               & 26                                                 & \textbf{128}       \\
Channel Loss [dB]                     & 24                                                  & 20                                                 & NaN                                                 & 18.4                                                & 32                                                 & 15        \\
Equalization                          & CTLE                                                & CTLE                                               & CTLE                                                & CTLE                                                & CTLE                                               & CTLE      \\
                                      & 4-tap DFE                                           & 2-tap DFE                                          & VGA                                                 & 1-tap DFE                                           & 4-tap DFE                                          & 1-tap DFE \\
Active~Area~[$\textrm{mm}^2$/lane]                & 0.065**                                              & 0.108                                              & 0.383                                               & 0.141                                               & 0.073                                              & 0.105     \\
RX Clocking Power Efficiency [mW/Gb/s] & 1.24**                                               & 0.51                                               & 0.92                                                & N/A                                                 & 0.73                                               & \textbf{0.65}      \\
Energy Efficiency [pJ/bit]            & 5.9**                                                & 2.02                                               & 2.59                                                & 2.53                                                & 3.3                                                & \textbf{1.81}      \\
{* CML PI ** Transceiver}                          &                                                     &                                                    &                                                     &                                                     &                                                    &           
\end{tblr}
\label{tab: 2}
\end{table*}

\subsection{RX Performance}
\begin{figure}
    \centering
    \includegraphics[width=1\linewidth]{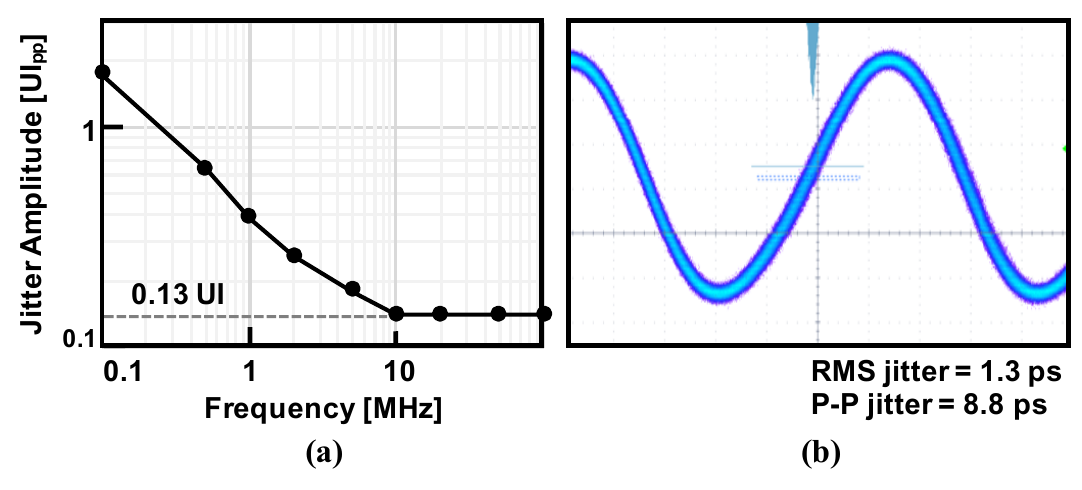}
    \caption{Measured (a) jitter tolerance and (b) recovered clock of the proposed CDR.}
    \label{fig: 24}
\end{figure}

\begin{figure}
    \centering
    \includegraphics[width=1\linewidth]{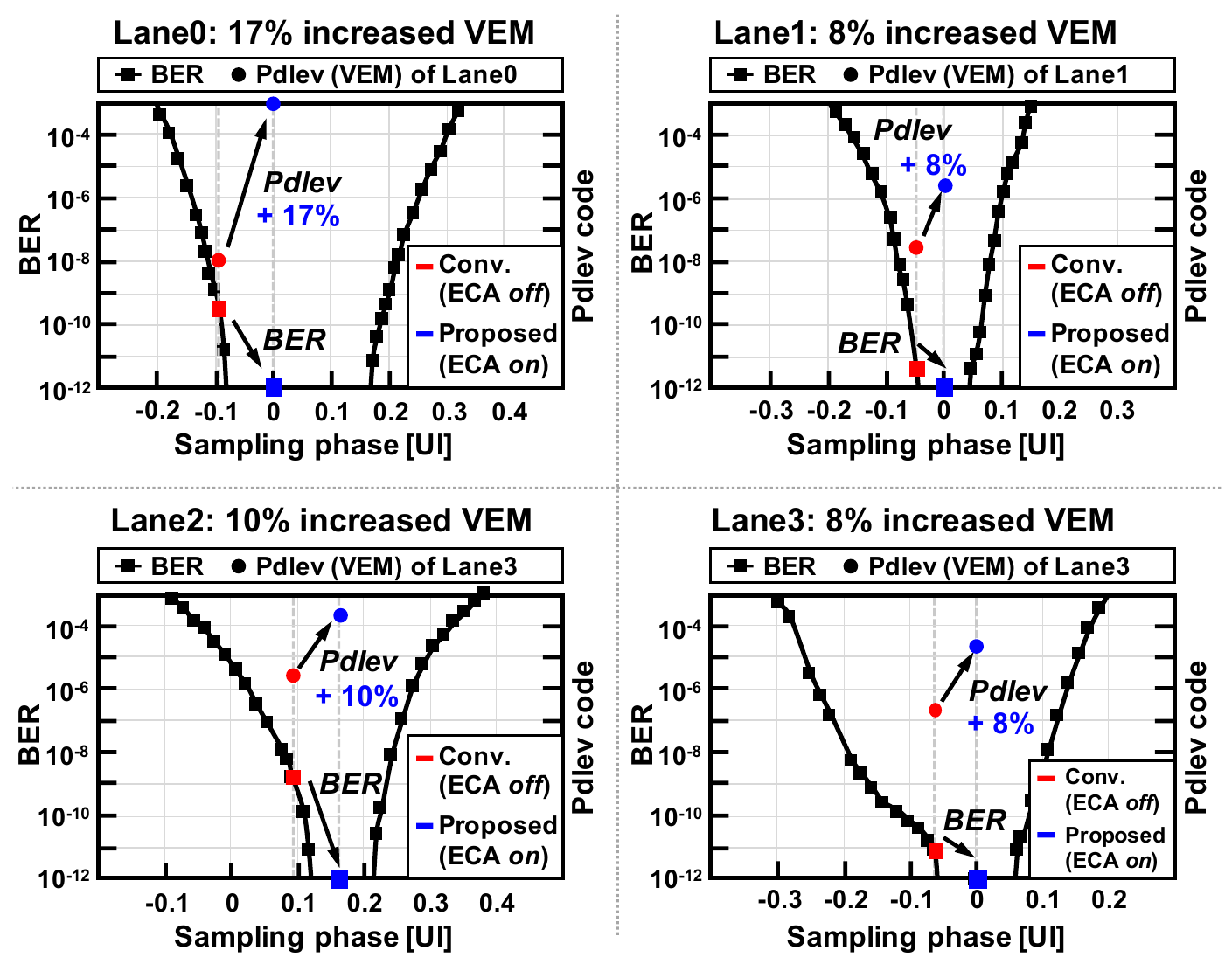}
    \caption{Measured BER bathtub curves of each lane.}
    \label{fig: 23}
\end{figure}

The RX performance at 32\,Gb/s was measured with a 15\,dB channel loss including SMA cable and FR-4 trace loss. 
Fig.~\ref{fig: 22} depicts the tested channel characteristic. 
As shown in Fig.~\ref{fig: 24}, the measured JTOL corner frequency is 10\,MHz, and the measured recovered clock jitter is 1.3\,$\textrm{ps}_\textrm{rms}$ and 8.8\,$\textrm{ps}_\textrm{p2p}$, respectively. 
Fig.~\ref{fig: 23} shows the measured BER bathtub curves for each lane of the proposed 4-lane RX. 
The effectiveness of the proposed ECA is validated by comparing BER when ECA is on/off. 
For lane0, with the conventional timing recovery (ECA off), the recovered clock phase locks at -0.09\,UI apart from the proposed CDR's sampling phase, which results in the degraded BER of $3\times10^{-10}$. 
On the other hand, the proposed CDR shifts the lock point with ECA, achieving error-free operation (BER $<10^{-12}$) and a 17\,\% increase in VEM compared to that of the conventional CDR. 
Across all lanes, BER improves from $3\times10^{-10}, 4\times10^{-12}, 1.5\times10^{-9}$, and $8\times10^{-12}$ in conventional CDR to BER $<10^{-12}$ with the proposed method. 
In addition, VEMs of four lanes are increased by 17\,\%, 8\,\%, 10\,\%, 8\,\%, respectively, thanks to ECA. 
In Table~\ref{tab: 2}, performance comparison with other recently published high-speed NRZ RXs is provided. 
Although other works implements fewer (one or two) lanes and a shorter-distance clock distribution, thanks to the novel low-power global clock distribution technique, the proposed 4-lane RX achieves the best energy efficiency.

\section{Conclusion} \label{section6}
In multi-lane high-speed RXs, the clocking scheme plays a crucial role in determining overall RX power consumption. 
This paper presents design techniques for an energy-efficient multi-lane baud-rate CDR. 
The proposed CDR achieves substantial power savings with an energy-efficient clocking scheme, employing a low-frequency single-phase global clock distribution and baud-rate recovery. 
Frequency offset is compensated by FDIV before global clock distribution, 
which leads to reduced deterministic jitter in the recovered clock typically caused by PI non-linearity.
In addition, the proposed background ECA successfully addresses the lock point issue of the conventional BRPDs, thereby improving both VEM and BER. 
This enhancement is achieved without requiring additional hardware. 
The prototype 4$\times$32Gb/s RX fabricated in a 28\,nm CMOS process demonstrates superior energy efficiency of 1.8\,pJ/bit even with a long-distance clock distribution of 2\,mm. 
Thanks to the frequency-tracking FDIV, the RX achieves a low integrated fractional spur of -40.4\,dBc with a 2500\,ppm frequency offset. 
Furthermore, the proposed ECA contributes to 17\,\% increase in VEM compared to conventional methods, ensuring robust and error-free operation of the proposed RX.


%

\section*{Acknowledgment}
The EDA tool was supported by the IC Design Education Center (IDEC), Korea.

\ifCLASSOPTIONcaptionsoff
  \newpage
\fi



\bibliographystyle{IEEEtran}
\bibliography{ref}

\begin{thebibliography}{10}
\providecommand{\url}[1]{#1}
\csname url@samestyle\endcsname
\providecommand{\newblock}{\relax}
\providecommand{\bibinfo}[2]{#2}
\providecommand{\BIBentrySTDinterwordspacing}{\spaceskip=0pt\relax}
\providecommand{\BIBentryALTinterwordstretchfactor}{4}
\providecommand{\BIBentryALTinterwordspacing}{\spaceskip=\fontdimen2\font plus
\BIBentryALTinterwordstretchfactor\fontdimen3\font minus \fontdimen4\font\relax}
\providecommand{\BIBforeignlanguage}[2]{{%
\expandafter\ifx\csname l@#1\endcsname\relax
\typeout{** WARNING: IEEEtran.bst: No hyphenation pattern has been}%
\typeout{** loaded for the language `#1'. Using the pattern for}%
\typeout{** the default language instead.}%
\else
\language=\csname l@#1\endcsname
\fi
#2}}
\providecommand{\BIBdecl}{\relax}
\BIBdecl

\bibitem{1}
B.~Zhang \emph{et~al.}, ``A 112{Gb/s} serial link transceiver with 3-tap {FFE} and 18-tap {DFE} receiver for up to 43d{B} insertion loss channel in 7nm {FinFET} technology,'' in \emph{2023 IEEE International Solid-State Circuits Conference (ISSCC)}.\hskip 1em plus 0.5em minus 0.4em\relax IEEE, 2023, pp. 5--7.

\bibitem{2}
N.~Kocaman \emph{et~al.}, ``An 182{mW} 1-60{Gb/s} configurable {PAM-4/NRZ} transceiver for large scale {ASIC} integration in 7nm {FinFET} technology,'' in \emph{2022 IEEE International Solid-State Circuits Conference (ISSCC)}, vol.~65.\hskip 1em plus 0.5em minus 0.4em\relax IEEE, 2022, pp. 120--122.

\bibitem{3}
M.-A. LaCroix \emph{et~al.}, ``A 116{Gb/s} {DSP}-based wireline transceiver in 7nm {CMOS} achieving 6{pJ/b} at 45{dB} loss in {PAM-4/Duo-PAM-4} and 52{dB} in {PAM-2},'' in \emph{2021 IEEE International Solid-State Circuits Conference (ISSCC)}, vol.~64.\hskip 1em plus 0.5em minus 0.4em\relax IEEE, 2021, pp. 132--134.

\bibitem{4}
B.-J. Yoo \emph{et~al.}, ``A 56{Gb/s} 7.7 {mW/Gb/s} {PAM-4} wireline transceiver in 10nm {FinFET} using {MM-CDR-Based} {ADC} timing skew control and low-power {DSP} with approximate multiplier,'' in \emph{2020 IEEE International Solid-State Circuits Conference-(ISSCC)}.\hskip 1em plus 0.5em minus 0.4em\relax IEEE, 2020, pp. 122--124.

\bibitem{5}
N.~Kocaman \emph{et~al.}, ``A 3.8 {mW/Gbps} quad-channel 8.5--13 {Gbps} serial link with a 5 tap {DFE} and a 4 tap transmit {FFE} in 28 nm cmos,'' \emph{IEEE Journal of Solid-State Circuits}, vol.~51, no.~4, pp. 881--892, 2016.

\bibitem{6}
B.~Raghavan \emph{et~al.}, ``A sub-2{W} 39.8-to-44.6{Gb/s} transmitter and receiver chipset with {SFI-5.2} interface in 40nm {CMOS},'' in \emph{2013 IEEE International Solid-State Circuits Conference-(ISSCC)}.\hskip 1em plus 0.5em minus 0.4em\relax IEEE, 2013, pp. 32--33.

\bibitem{7}
U.~Singh \emph{et~al.}, ``A 780{mW} 4×28{Gb/s} transceiver for 100{GbE} {G}earbox {PHY} in 40nm {CMOS},'' in \emph{2014 IEEE International Solid-State Circuits Conference-(ISSCC)}.\hskip 1em plus 0.5em minus 0.4em\relax IEEE, 2014, pp. 40--41.

\bibitem{low_power}
J.~Pangjun and S.~S. Sapatnekar, ``Low-power clock distribution using multiple voltages and reduced swings,'' \emph{IEEE Transactions on Very Large Scale Integration (VLSI) Systems}, vol.~10, no.~3, pp. 309--318, 2002.

\bibitem{casper2009clocking}
B.~Casper and F.~O'Mahony, ``Clocking analysis, implementation and measurement techniques for high-speed data links—a tutorial,'' \emph{IEEE Transactions on Circuits and Systems I: Regular Papers}, vol.~56, no.~1, pp. 17--39, 2009.

\bibitem{8}
S.~Saxena, G.~Shu, R.~K. Nandwana, M.~Talegaonkar, A.~Elkholy, T.~Anand, W.-S. Choi, and P.~K. Hanumolu, ``A 2.8 {mW/Gb/s}, 14 {Gb/s} serial link transceiver,'' \emph{IEEE Journal of Solid-State Circuits}, vol.~52, no.~5, pp. 1399--1411, 2017.

\bibitem{9}
S.~Park, Y.~Choi, J.~Sim, J.~Choi, H.~Park, Y.~Kwon, and C.~Kim, ``A 0.83 {pJ/b} 52{Gb/s} {PAM-4} baud-rate {CDR} with pattern-based phase detector for short-reach applications,'' in \emph{2023 IEEE International Solid-State Circuits Conference (ISSCC)}.\hskip 1em plus 0.5em minus 0.4em\relax IEEE, 2023, pp. 118--120.

\bibitem{10}
Y.~Segal, A.~Laufer, A.~Khairi, Y.~Krupnik, M.~Cusmai, I.~Levin, A.~Gordon, Y.~Sabag, V.~Rahinski, G.~Ori \emph{et~al.}, ``A 1.41 {pJ/b} 224{Gb/s} {PAM-4} {SerDes} receiver with 31{dB} loss compensation,'' in \emph{2022 IEEE International Solid-State Circuits Conference (ISSCC)}, vol.~65.\hskip 1em plus 0.5em minus 0.4em\relax IEEE, 2022, pp. 114--116.

\bibitem{11}
B.~Ye, K.~Sheng, W.~Gai, H.~Niu, B.~Zhang, Y.~He, S.~Jia, C.~Chen, and J.~Yu, ``A 2.29 {pJ/b} 112{Gb/s} wireline transceiver with {RX} 4-tap {FFE} for medium-reach applications in 28nm {CMOS},'' in \emph{2022 IEEE International Solid-State Circuits Conference (ISSCC)}, vol.~65.\hskip 1em plus 0.5em minus 0.4em\relax IEEE, 2022, pp. 118--120.

\bibitem{12}
J.~Bailey, H.~Shakiba, E.~Nir, G.~Marderfeld, P.~Krotnev, M.-A. LaCroix, and D.~Cassan, ``A 112{Gb/s} {PAM-4} low-power 9-tap sliding-block {DFE} in a 7nm {FinFET} wireline receiver,'' in \emph{2021 IEEE International Solid-State Circuits Conference (ISSCC)}, vol.~64.\hskip 1em plus 0.5em minus 0.4em\relax IEEE, 2021, pp. 140--142.

\bibitem{18}
F.~Spagna, L.~Chen, M.~Deshpande, Y.~Fan, D.~Gambetta, S.~Gowder, S.~Iyer, R.~Kumar, P.~Kwok, R.~Krishnamurthy \emph{et~al.}, ``A 78{mW} 11.8 {Gb/s} serial link transceiver with adaptive {RX} equalization and baud-rate {CDR} in 32nm {CMOS},'' in \emph{2010 IEEE International Solid-State Circuits Conference-(ISSCC)}.\hskip 1em plus 0.5em minus 0.4em\relax IEEE, 2010, pp. 366--367.

\bibitem{19}
R.~Dokania, A.~Kern, M.~He, A.~Faust, R.~Tseng, S.~Weaver, K.~Yu, C.~Bil, T.~Liang, and F.~O'Mahony, ``A 5.9 {pJ/b} 10{Gb/s} serial link with unequalized {MM-CDR} in 14nm tri-gate {CMOS},'' in \emph{2015 IEEE International Solid-State Circuits Conference-(ISSCC) Digest of Technical Papers}.\hskip 1em plus 0.5em minus 0.4em\relax IEEE, 2015, pp. 1--3.

\bibitem{20}
M.-C. Choi, H.-G. Ko, J.~Oh, H.-Y. Joo, K.~Lee, and D.-K. Jeong, ``A 0.1-{pJ/b/dB} 28-{Gb/s} maximum-eye tracking, weight-adjusting {MM CDR} and adaptive {DFE} with single shared error sampler,'' in \emph{2020 IEEE Symposium on VLSI Circuits}.\hskip 1em plus 0.5em minus 0.4em\relax IEEE, 2020, pp. 1--2.

\bibitem{13}
J.~F. Bulzacchelli, M.~Meghelli, S.~V. Rylov, W.~Rhee, A.~V. Rylyakov, H.~A. Ainspan, B.~D. Parker, M.~P. Beakes, A.~Chung, T.~J. Beukema \emph{et~al.}, ``A 10-{Gb/s} 5-tap {DFE}/4-tap {FFE} transceiver in 90-nm {CMOS} technology,'' \emph{IEEE Journal of Solid-State Circuits}, vol.~41, no.~12, pp. 2885--2900, 2006.

\bibitem{15}
P.~K. Hanumolu, G.-Y. Wei, and U.-K. Moon, ``A wide-tracking range clock and data recovery circuit,'' \emph{IEEE Journal of Solid-State Circuits}, vol.~43, no.~2, pp. 425--439, 2008.

\bibitem{rambus}
M.~Hossain, E.-H. Chen, R.~Navid, B.~Leibowitz, A.~Chou, S.~Li, M.-J. Park, J.~Ren, B.~Daly, B.~Su \emph{et~al.}, ``A 4$\times$ 40 gb/s quad-lane cdr with shared frequency tracking and data dependent jitter filtering,'' in \emph{2014 Symposium on VLSI Circuits Digest of Technical Papers}.\hskip 1em plus 0.5em minus 0.4em\relax IEEE, 2014, pp. 1--2.

\bibitem{16}
R.~K. Nandwana, S.~Saxena, A.~Elkholy, M.~Talegaonkar, J.~Zhu, W.-S. Choi, A.~Elmallah, and P.~K. Hanumolu, ``A 3-to-10{Gb/s} 5.75 {pJ/b} transceiver with flexible clocking in 65nm {CMOS},'' in \emph{2017 IEEE International Solid-State Circuits Conference (ISSCC)}.\hskip 1em plus 0.5em minus 0.4em\relax IEEE, 2017, pp. 492--493.

\bibitem{17}
A.~Agrawal, A.~Liu, P.~K. Hanumolu, and G.-Y. Wei, ``An 8$\times$5 {Gb/s} parallel receiver with collaborative timing recovery,'' \emph{IEEE journal of solid-state circuits}, vol.~44, no.~11, pp. 3120--3130, 2009.

\bibitem{PR1}
S.~Chen, L.~Zhou, I.~Zhuang, J.~Im, D.~Melek, J.~Namkoong, M.~Raj, J.~Shin, Y.~Frans, and K.~Chang, ``A 4-to-16ghz inverter-based injection-locked quadrature clock generator with phase interpolators for multi-standard i/os in 7nm finfet,'' in \emph{2018 IEEE International Solid-State Circuits Conference - (ISSCC)}, 2018, pp. 390--392.

\bibitem{PR2}
E.~Monaco, G.~Anzalone, G.~Albasini, S.~Erba, M.~Bassi, and A.~Mazzanti, ``A 2–11 ghz 7-bit high-linearity phase rotator based on wideband injection-locking multi-phase generation for high-speed serial links in 28-nm cmos fdsoi,'' \emph{IEEE Journal of Solid-State Circuits}, vol.~52, no.~7, pp. 1739--1752, 2017.

\bibitem{8b}
Y.-C. Huang and B.-J. Chen, ``An 8b injection-locked phase rotator with dynamic multiphase injection for 28/56/112{Gb/s} serdes application,'' in \emph{2019 IEEE International Solid-State Circuits Conference-(ISSCC)}.\hskip 1em plus 0.5em minus 0.4em\relax IEEE, 2019, pp. 486--488.

\bibitem{7b}
Z.~Wang, Y.~Zhang, Y.~Onizuka, and P.~R. Kinget, ``A high-accuracy multi-phase injection-locked 8-phase 7{GHz} clock generator in 65nm with 7b phase interpolators for high-speed data links,'' in \emph{2021 IEEE International Solid-State Circuits Conference (ISSCC)}, vol.~64.\hskip 1em plus 0.5em minus 0.4em\relax IEEE, 2021, pp. 186--188.

\bibitem{7027236}
A.~Elkholy, T.~Anand, W.-S. Choi, A.~Elshazly, and P.~K. Hanumolu, ``A 3.7 m{W} low-noise wide-bandwidth 4.5 {GH}z digital fractional-{N} {PLL} using time amplifier-based {TDC},'' \emph{IEEE Journal of Solid-State Circuits}, vol.~50, no.~4, pp. 867--881, 2015.

\bibitem{MET}
H.-Y. Joo, J.~Lee, H.~Ju, H.-G. Ko, J.~M. Yoon, B.~Kang, and D.-K. Jeong, ``A maximum-eye-tracking {CDR} with biased data-level and eye slope detector for near-optimal timing adaptation,'' \emph{IEEE Transactions on Very Large Scale Integration (VLSI) Systems}, vol.~28, no.~12, pp. 2708--2720, 2020.

\bibitem{56g}
T.~Shibasaki, W.~Chaivipas, Y.~Chen, Y.~Doi, T.~Hamada, H.~Takauchi, T.~Mori, Y.~Koyanagi, and H.~Tamura, ``A 56-{Gb/s} receiver front-end with a {CTLE} and 1-tap {DFE} in 20-nm {CMOS},'' in \emph{2014 Symposium on VLSI Circuits Digest of Technical Papers}.\hskip 1em plus 0.5em minus 0.4em\relax IEEE, 2014, pp. 1--2.

\bibitem{28}
T.~Shibasaki, T.~Danjo, Y.~Ogata, Y.~Sakai, H.~Miyaoka, F.~Terasawa, M.~Kudo, H.~Kano, A.~Matsuda, S.~Kawai \emph{et~al.}, ``A 56{Gb/s} {NRZ}-electrical 247{mW}/lane serial-link transceiver in 28nm {CMOS},'' in \emph{2016 IEEE International Solid-State Circuits Conference (ISSCC)}.\hskip 1em plus 0.5em minus 0.4em\relax IEEE, 2016, pp. 64--65.

\bibitem{23}
B.~Razavi, ``The design of an equalizer—part one [the analog mind],'' \emph{IEEE Solid-State Circuits Magazine}, vol.~13, no.~4, pp. 7--160, 2021.

\bibitem{24}
------, ``The decision-feedback equalizer [a circuit for all seasons],'' \emph{IEEE Solid-State Circuits Magazine}, vol.~9, no.~4, pp. 13--132, 2017.

\bibitem{25}
------, ``The design of a comparator [the analog mind],'' \emph{IEEE Solid-State Circuits Magazine}, vol.~12, no.~4, pp. 8--14, 2020.

\bibitem{21}
A.~Elkholy, S.~Saxena, R.~K. Nandwana, A.~Elshazly, and P.~K. Hanumolu, ``A 2.0--5.5 {GHz} wide bandwidth ring-based digital fractional-n pll with extended range multi-modulus divider,'' \emph{IEEE Journal of Solid-State Circuits}, vol.~51, no.~8, pp. 1771--1784, 2016.

\bibitem{8357042}
A.~Elmallah, M.~G. Ahmed, A.~Elkholy, W.-S. Choi, and P.~K. Hanumolu, ``A 1.6ps peak-{INL} 5.3ns range two-step digital-to-time converter in 65nm {CMOS},'' in \emph{2018 IEEE Custom Integrated Circuits Conference (CICC)}, 2018, pp. 1--4.

\bibitem{22}
A.~Elkholy, M.~Talegaonkar, T.~Anand, and P.~K. Hanumolu, ``Design and analysis of low-power high-frequency robust sub-harmonic injection-locked clock multipliers,'' \emph{IEEE Journal of Solid-State Circuits}, vol.~50, no.~12, pp. 3160--3174, 2015.

\bibitem{7999180}
M.~Talegaonkar, T.~Anand, A.~Elkholy, A.~Elshazly, R.~K. Nandwana, S.~Saxena, B.~Young, W.-S. Choi, and P.~K. Hanumolu, ``A 5{GH}z digital fractional-{$N$} {PLL} using a 1-bit delta–sigma frequency-to-digital converter in 65 nm {CMOS},'' \emph{IEEE Journal of Solid-State Circuits}, vol.~52, no.~9, pp. 2306--2320, 2017.

\bibitem{29}
Y.-C. Liu, W.-Z. Chen, Y.-S. Lee, Y.-H. Chen, S.~Ming, and Y.-H. Lin, ``A 103 {fJ/b/dB}, 10--26 {Gb/s} receiver with a dual feedback nested loop cdr for wide bandwidth jitter tolerance enhancement,'' \emph{IEEE Journal of Solid-State Circuits}, 2023.

\end{thebibliography}
%

%




\end{document}